\documentclass[sn-mathphys-num]{sn-jnl}

\usepackage{graphicx}%
\usepackage{multirow}%
\usepackage{amsmath,amssymb,amsfonts}%
\usepackage{amsthm}%
\usepackage{mathrsfs}%
\usepackage[title]{appendix}%
\usepackage{xcolor}%
\usepackage{textcomp}%
\usepackage{manyfoot}%
\usepackage{booktabs}%
\usepackage{algorithm}%
\usepackage{algorithmicx}%
\usepackage{algpseudocode}%
\usepackage{listings}%
\usepackage{braket}

\newcommand{\tr}{\mathrm{tr}\,}

\newcommand{\Ad}{\mathrm{Ad}}
\newcommand{\ad}{\mathrm{ad}}

\newcommand{\Exp}{\mathrm{Exp}_{\star}}

\theoremstyle{thmstyleone}%
\theoremstyle{thmstyletwo}%

\theoremstyle{thmstylethree}%

\begin{document}

\title{Star product for qubit states in phase space and star exponentials}


\author*[1]{\fnm{Jasel} \sur{ Berra--Montiel}}\email{jasel.berra@uaslp.mx}
\equalcont{These authors contributed equally to this work.}

\author[1]{\fnm{Alberto} \sur{Molgado}}\email{alberto.molgado@uaslp.mx}
\equalcont{These authors contributed equally to this work.}

\author[1]{\fnm{Mar} \sur{S\'anchez--C\'ordova}}\email{maria.cordova@uaslp.mx}
\equalcont{These authors contributed equally to this work.}

\affil*[1]{\orgdiv{Facultad de Ciencias}, \orgname{Universidad Aut\'onoma de San Luis 
	Potos\'{\i}}, \orgaddress{\street{Campus Pedregal, Av. Parque Chapultepec 1610, Col. Privadas del Pedregal}, \city{San
	Luis Potos\'{\i}}, \postcode{78217}, \state{SLP}, \country{M\'exico}}}


\abstract{In this paper,  we formulate the phase space description of qubit systems using coadjoint orbits of $SU(2)$ and the Stratonovich–Weyl correspondence, yielding a deformation quantization on the sphere. The resulting star product reproduces the operator algebra of complexified quaternions and its antisymmetric part induces the Lie–Poisson structure associated with the Kirillov–Kostant–Souriau symplectic form. We show that quantum dynamics can be expressed entirely in phase space through star exponentials of Hamiltonian symbols, leading to an explicit representation of the propagator. Further, we establish the equivalence between the coherent-state path integral formulation on $S^{2}$ and the algebraic description in terms of star exponentials. Some examples illustrating the construction of the star-exponential functions and the resulting Poisson structure are included.}




\maketitle
\section{Introduction}
The formulation of quantum mechanics on phase space provides a conceptually and technically powerful framework in which classical and quantum structures can be treated on equal footing \cite{Curtright, Overview}. In this approach, quantum observables are represented by functions on a classical phase space endowed with a noncommutative associative product, the so-called star product, which encodes the operator algebra. This representation, formalized within deformation quantization program, replaces the canonical commutation relations by a deformation of the commutative algebra $C^{\infty}(M)$ of smooth functions on a symplectic manifold $M$ such that the leading order of the antisymmetric part of the star product reproduces the Poisson bracket \cite{Bayen}. While deformation quantization is well understood for continuous varaibles quantum systems on flat phase spaces, $M=\mathbb{R}^{2n}$, where the Moyal product provides an explicit realization of the star product, recent developments have shown that these methods may be successfully extended to either curved or dynamical spacetimes, including cosmological backgrounds and black-hole geometries. In such frameworks, phase-space techniques offer a natural language for describing quantum dynamics in gravitational environments \cite{Bobadilla, Cembranos}. In contrast, the extension to finite-dimensional quantum systems requires a different geometric setting \cite{Varilly,Heiss, Tilma, Klimov, Klimov2}. In particular, for spin systems, the algebra of observables is finite-dimensional and non-abelian, and no global system of canonical coordinates exists. Instead, a natural phase space is provided by the coadjoint orbits of the underlying symmetry group. In the case of a qubit, the relevant group is $SU(2)$, and its nontrivial coadjoint orbits are two-dimensional spheres $S^{2}$, which carry a canonical symplectic structure given by the Kirillov–Kostant–Souriau (KKS) form \cite{GQ,Orbit}. This construction equips $S^{2}$ with a Poisson bracket inherited from the Lie algebra structure on $\mathfrak{su}(2)^{*}$, thereby providing a natural classical phase space for spin systems.

Quantization on such curved phase spaces can be implemented through the Stratonovich–Weyl (SW) correspondence, which establishes an equivariant isomorphism between operators acting on a finite-dimensional Hilbert space and functions on the coadjoint orbit. This correspondence allows one to define phase-space symbols and induces, via operator multiplication, an associative star product on $C^{\infty}(S^{2})$. In contrast to the formal deformation approach where the star product is defined as an asymptotic formal power series in $\hbar$  without guaranteed convergence \cite{Bayen,Kontsevich}, the resulting star product on the sphere is exact and finite, reflecting the underlying matrix algebra structure. Its antisymmetric part defines a Moyal bracket which reproduces the $\mathfrak{su}(2)$ Lie algebra and, upon suitable identification, induces the Poisson structure associated with the KKS symplectic form.

The aim of this work is to explicitly construct and analyze the star product associated with the SW correspondence for qubit systems and, also,  to investigate its role in the phase-space formulation of quantum dynamics. We show that the star product provides a faithful realization of the operator algebra on $M_{2}(\mathbb{C})$ in terms of functions on the sphere, and, besides that such a star product captures the full noncommutative structure of theory. Furthermore, it can be naturally interpreted as a phase-space realization
of the algebra of complexified quaternions. In addition, we develop the notion of star exponentials for Hamiltonian symbols on $S^{2}$ and demonstrate that they yield a natural phase-space representation of the unitary time-evolution operator. In this way, the propagator can be expressed entirely in terms of phase-space quantities, leading to an exact formulation of quantum dynamics within the deformation quantization framework. In parallel, we also consider the path integral representation of the propagator in terms of $SU(2)$ coherent states, which defines a functional integral over trajectories on the sphere. We show that these two descriptions, the algebraic formulation in terms of star exponentials and the geometric formulation in terms of path integrals, are equivalent and constitute two complementary realizations of quantum dynamics on curved phase space. This correspondence extends the well-known equivalence between Moyal quantization and Feynman path integrals in $\mathbb{R}^{2n}$ to the setting of coadjoint orbits.

The paper is organized as follows. In Section~\ref{sec:Orbits}, we review the construction of the coadjoint orbits of the Lie group $SU(2)$ together with their symplectic structure. Section~\ref{sec:SW} is devoted to the construction of a star product on this symplectic manifold via the Stratonovich–Weyl correspondence. In Section~\ref{sec:SE}, we analyze the relation between star-exponential functions and quantum mechanical propagators, and discuss their natural connection within the framework of Feynman’s path integral. We also include explicit examples illustrating the construction of the star-exponential functions and the resulting Poisson structure. Finally, we conclude in Section~\ref{sec:Conclusions} with a summary and outlook.

\section{Coadjoint orbits and the phase space of qubits}
\label{sec:Orbits}

In contrast to continuous-variable quantum systems, where phase space is identified with the symplectic manifold $\mathbb{R}^{2n}$, endowed with its standard Poisson structure, finite-dimensional quantum systems usually do not admit a globally defined phase space of this form. In particular, for spin systems, the algebra of observables is finite-dimensional and non-abelian, preventing the use of global canonical position–momentum coordinates. However, for finite dimensional quantum system a natural notion of phase space is instead obtained via the coadjoint orbit method, in which a symplectic manifold structure is obtained by analyzing the orbits of the coadjoint action of a Lie group on the dual of its Lie algebra  \cite{Kostant, Kirillov}. In the case of qubits, this construction identifies the phase space with a two-dimensional coadjoint orbit of $SU(2)$, equipped with the so called Kirillov–Kostant–Souriau symplectic form, thereby providing a well-defined Poisson structure and, as a consequence, a natural setting for deformation quantization. In this section, we closely follow Refs. \cite{Vogan,Hall} for notation and for standard results concerning the group $SU(2)$.

To make this identification explicit, let $\mathfrak{su}(2)$ be the Lie algebra of the group $SU(2)$, generated by
\begin{equation}
    J_{i}=\frac{1}{2}\sigma_{i}, \;\; i=1,2,3,
\end{equation}
where $\sigma_{i}$ denotes the Pauli matrices. With this normalization, the generators satisfy the commutation relations
\begin{equation}
    [J_{i},J_{j}]=i\epsilon_{ijk}J_{k},
\end{equation}
and provide the fundamental spin-1/2 representation of the $\mathfrak{su}(2)$ Lie algebra. As a real algebra, $\mathfrak{su}(2)$ is three dimensional and simple, which means that any element $X\in \mathfrak{su}(2)$ can be written uniquely as
\begin{equation}
    X=x_{i}J_{i},
\end{equation}
where $x_{i}\in\mathbb{R}$ and the Einstein summation convention over repeated indices is understood. Moreover, the Killing form on $\mathfrak{su}(2)$, given by
\begin{equation}\label{Killing}
\kappa(X,Y)=-2\tr(X,Y)    
\end{equation}
is non-degenerate and therefore induces a canonical identification between the Lie algebra and its dual, $\mathfrak{su}(2)\simeq \mathfrak{su}(2)^{*}$, via the correspondence $X\mapsto \kappa(X,\cdot)$.

\noindent The adjoint action of $SU(2)$ on $\mathfrak{su}(2)$ is the smooth group homomorphism 
\begin{equation}
    \Ad:SU(2)\to \mathrm{Aut(}\mathfrak{su}(2)), 
\end{equation}
defined by
\begin{equation}
    \Ad_{g}(X):=\frac{d}{dt}g\exp(tX)g^{-1}\Big|_{t=0}, \;\;g\in SU(2), X\in\mathfrak{su}(2).
\end{equation}
where $g\in SU(2)$, and $X\in\mathfrak{su}(2)$. Since $SU(2)$ corresponds to a matrix Lie group, the adjoint action takes the explicit form
\begin{equation}
    \Ad_{g}(X)=gXg^{-1}.
\end{equation}
Moreover, the adjoint action is a Lie algebra homomorphism from $\mathfrak{su}(2)$ into itself, since it preserves the Lie braket
\begin{equation}
    \Ad_{g}[X,Y]=[\Ad_{g}X,\Ad_{g}Y].
\end{equation}
The infinitesimal version of the adjoint action is the adjoint representation
\begin{equation}
    \ad:\mathfrak{su}(2)\to \mathrm{End}(\mathfrak{su}(2)),
\end{equation}
given by
\begin{equation}
    \ad_{X}(Y):=[X,Y], \;\; X,Y\in\mathfrak{su}(2).
\end{equation}
This map is obtained by differentiating the adjoint action at the identity element of the group. More precisely, for any $X,Y\in\mathfrak{su}(2)$,
\begin{equation}
    \ad_{X}(Y)=\frac{d}{dt}\Ad _{\exp(tX)}(Y)\Big|_{t=0}.
\end{equation}
The adjoint representation defines a homomorphism of Lie algebras, in the sense that
\begin{equation}
    [\ad_{X},\ad_{Y}]=\ad_{[X,Y]}.
\end{equation}
Let $\mathfrak{su}(2)^{*}$ denote the dual Lie algebra of $\mathfrak{su}(2)$. The coadjoint action 
\begin{equation}
    \Ad^{*}:SU(2)\to \mathrm{Aut}(\mathfrak{su}(2)^{*}),
\end{equation}
is defined as the dual of the adjoint action
\begin{equation}
    \braket{\Ad_{g}^{*}\xi,X}:=\braket{\xi,\Ad_{g^{-1}}X},
\end{equation}
where $X\in \mathfrak{su}(2)$ and $\xi\in\mathfrak{su}(2)^{*}$. Its infinitesimal version reads
\begin{equation}
    (\ad_{X}^{*}\xi)(Y)=-\xi([X,Y]), \;\;X,Y\in\mathfrak{su}(2).
\end{equation}
The coadjoint orbit of $\xi\in\mathfrak{su}(2)$ is given by
\begin{equation}
    \mathcal{O}_{\xi}:=\left\{\Ad^{*}_{g}\xi:g\in SU(2)\right\}\subset \mathfrak{su}(2)^{*}.
\end{equation}
Coadjoint orbits are immersed homogeneous submanifolds of dual Lie algebras carrying canonical symplectic structures which play a fundamental role in Lie theory, symplectic geometry, and geometric quantization.

\noindent To make the construction of the orbits explicit, let us fix $\xi\in\mathfrak{su}(2)^{*}$. Then, since the Killing form (\ref{Killing}) is $\Ad$-invariant, namely
\begin{equation}
\kappa(\Ad_{g}X, \Ad_{g}Y)=\kappa(X,Y),
\end{equation}
the identification of $\mathfrak{su}(2)$ with $\mathfrak{su}(2)^{*}$ is also $SU(2)$-equivariant. More precisely, the linear map $\Phi:\mathfrak{su}(2)\to\mathfrak{su}(2)^{*}$ defined by $\Phi(X)=\kappa(X,\cdot)$ is not only an isomorphism of vector spaces, but also intertwines the adjoint and coadjoint actions, in the sense that $\Ad_{g}^{*}\Phi(X)=\Phi(\Ad_{g}X)$. As a consequence, under this identification the coadjoint and adjoint actions coincide. It then follows from the well-known fact that the adjoint representation of 
$SU(2)$ is isomorphic to the representation of $SO(3)$, the adjoint action may be written as $\Ad_{g}(X)=(R(g)\vec{x})_{i}J_{i}$, where $X=x_{i}J_{i}$ and $R(g)$ is a $3\times 3$ orthogonal matrix with unit determinant associated with $g\in SU(2)$. 

\noindent Since $R(g)$ is an orthogonal matrix, it preserves the Euclidean norm, and therefore 
\begin{equation}
    ||R(g)\vec{x}||^{2}=||\vec{x}||=\kappa(X,X)=||X||^{2}.
\end{equation}
Consequently, the adjoint orbit of $X$, $\mathcal{O}_{X}:=\left\{\Ad_{g}X:g\in SU(2)\right\}$ is given by
\begin{equation}
    \mathcal{O}_{X}=\left\{Y\in \mathfrak{su}(2):||Y||=||X||\right\}\simeq S^{2}_{||\vec{x}||}.
\end{equation}
By equivariance of the identification $\mathfrak{su}(2)\simeq\mathfrak{su}^{*}(2)$, adjoint and coadjoint orbits are mapped into each other orbit by orbit. We therefore conclude that the non trivial coadjoint orbits of $SU(2)$ are two dimensional spheres.

Finally, a symplectic structure can be constructed by considering the coadjoint orbits of a Lie group acting on the dual space of its Lie algebra \cite{GQ,Orbit}. For the adjoint orbit through a point $X\in\mathfrak{su}(2)$, the tangent space is given by 
\begin{equation}
    T_{X}\mathcal{O}_{X}:=\left\{\ad_{Y}X:Y\in\mathfrak{su}(2)\right\}.
\end{equation}
Using the $SU(2)$-equivariant identification between $\mathfrak{su}(2)$ and $\mathfrak{su}(2)^{*}$, an analogous expression holds for the coadjoint orbit through $\xi\in\mathfrak{su}(2)^{*}$, namely
\begin{equation}
    T_{\xi}\mathcal{O}_{\xi}:=\left\{\ad^{*}_{Y}\xi:Y\in\mathfrak{su}(2)\right\}.
\end{equation}
\noindent Every coadjoint orbit carries a canonical symplectic structure \cite{Vaisman}, known as Kirillov–Kostant–Souriau (KKS) symplectic form, defined at $\xi\in\mathcal{O}_{\xi}$ by 
\begin{equation}
    \omega_{\xi}(\ad^{*}_{X}\xi,\ad^{*}_{Y}\xi)=\braket{\xi,[X,Y]}, \,\, X,Y \in \mathfrak{su}(2). 
\end{equation}
The form $\omega_{\xi}$ is bilinear, antisymmetric, closed, and non-degenerate on the orbit, thereby providing the coadjoint orbit $\mathcal{O}_{\xi}$ with the structure of a symplectic manifold. In the $\mathfrak{su}(2)$ case, the KKS symplectic form admits the explicit expression
\begin{equation}\label{wKKS}
    \omega_{KKS}=\frac{1}{2||\vec{x}||^{2}}\epsilon_{ijk}x_{i}dx_{j}\wedge dx_{k},
\end{equation}
where $\vec{x}\in\mathbb{R}^{3}$ parametrized the orbit via the previous identification $\mathfrak{su}(2)\simeq\mathbb{R}^{3}$. This symplectic form is invariant under the $SU(2)$ action and coincides, up to normalization, with the standard area form on the two-sphere.

The Kirillov–Kostant–Souriau symplectic structure on coadjoint orbits provides the fundamental geometric input for both deformation quantization and geometric quantization of finite-dimensional quantum systems. On the one hand, the KKS symplectic form not only provides the measure on the classical phase space $C^{\infty}(\mathcal{O}_{\xi})$, but it also determines a canonical Poisson bracket, which serves as the semiclassical limit of the associative star products defined on the orbit. In the case of $SU(2)$, this Poisson structure underlies the construction of star products on the sphere $S^{2}$, including those relevant for spin and qubit systems, and fixes the normalization of the corresponding star exponentials through the orbit radius. On the other hand, from the perspective of geometric quantization, the integrability of the KKS symplectic form selects the admissible coadjoint orbits and establishes a correspondence between integral orbits and irreducible unitary representations of the group with the spin-$j$ representation, thereby providing a geometric interpretation of finite-dimensional Hilbert spaces.

\section{Phase-space quantization and the Stratonovich–Weyl map for qubits}\label{sec:SW}
Quantization on coadjoint orbits proceeds by associating classical observables, given by smooth functions $C^{\infty}(S^{2})$ on the orbit, with quantum operators acting on a Hilbert space. This association is implemented through the introduction of the Stratonovich-Weyl (SW) correspondence, which provides a consistent, covariant framework for mapping functions on the classical phase space to operators in the quantum theory. For a qubit system, the Hilbert space is $\mathbb{C}^{2}$ and quantum observables are therefore represented by Hermitian operators acting on this space, i.e., $2\times 2$ Hermitian matrices. The SW correspondence is realized through a family of operator-valued distributions $\hat{\Delta}(\Omega)$, referred to as the Stratonovich–Weyl kernel, where $\Omega$ labels points of the classical phase space, typically parametrized by coordinates on the two-sphere $S^{2}$. The kernel is assumed to satisfy the following physically motivated postulates \cite{Stratonovich}: 
\begin{enumerate}
	\item The mapping $W_{\hat{A}}(\Omega)=\tr\left[ \hat{A}\hat{\Delta}(\Omega)\right] $ defines a one-to-one linear correspondence, where $W_{\hat{A}}(\Omega)$ denotes the phase-space symbol associated with the operator $\hat{A}$. The inverse map is given by $\hat{A}=\int_{\Omega}W_{\hat{A}}(\Omega)\hat{\Delta}(\Omega)d\Omega$, so that the operator $\hat{A}$ can be uniquely reconstructed from its phase-space representation and viceversa.
	\item The phase-space function $W_{\hat{A}}(\Omega)$ is real-valued, which implies that the kernel $\hat{\Delta}(\Omega)$ must be Hermitian.
	\item The function $W_{\hat{A}}(\Omega)$ satisfies the standarization condition, namely that its integral over the entire phase space exists and reproduces the trace of the operator, $\int_{\Omega}W_{\hat{A}}(\Omega)d\Omega=\tr \hat{A}$, which is equivalently expressed by the normalization condition on the kernel $\int_{\Omega}\hat{\Delta}(\Omega)d\Omega=\hat{1}$.
	\item The SW correspondence satisfies the traciality condition, $\int_{\Omega}W_{\hat{A}}(\Omega)W_{\hat{B}}(\Omega)d\Omega=\tr\left[\hat{A}\hat{B} \right] $. This condition guarantees that the Hilbert–Schmidt inner product of operators is faithfully reproduced by the phase-space representation.
	\item The SW correspondence satisfies the covariance condition, which guarantees compatibility with the action of the symmetry group. Specifically, for any $g\in SU(2)$,
    \begin{equation}
        \hat{\Delta}(g\cdot\Omega)=U(g)\hat{\Delta}(\Omega)U(g)^{\dagger},
    \end{equation}
where $g\cdot\Omega$ stands for the natural action of $SU(2)$ on the phase space, while $U(g)$ denotes the unitary representation of $SU(2)$ on the Hilbert space. Consequently, the phase-space functions transform covariantly,
\begin{equation}
    W_{U(g)\hat{A}(\Omega)U(g)^{\dagger}}(\Omega)=W_{\hat{A}}(g^{-1}\cdot\Omega).
\end{equation}
In the spin-1/2 case, where the phase space is the sphere $S^{2}$, this action corresponds to spatial rotations on the sphere.
\end{enumerate}  

For the case of a two-level system, such as a qubit represented over the Bloch sphere, the SW kernel takes the form \cite{Varilly, Tilma}, 
\begin{eqnarray}\label{Delta}
	\hat{\Delta}_q(\theta,\phi)&=&\frac{1}{2}\hat{U}(\phi,\theta,\Phi)\hat{\Pi}_q\hat{U}(\phi,\theta,\Phi)^{\dagger}, \nonumber\\
	&=&\frac{1}{2}\left(
	\begin{array}{cc}
		1+\sqrt{3}\cos{\theta} & \sqrt{3}e^{-i\phi}\sin{\theta} \\
		\sqrt{3}e^{i\phi}\sin{\theta} & 1-\sqrt{3}\cos{\theta} 
	\end{array}
	\right).
\end{eqnarray}
where $0\leq\theta\leq \pi$, $0\leq\phi\leq 2\pi$ and $0\leq\Phi\leq 4\pi$ denote the conventional Euler angles parametrizing the $SU(2)$ rotations \cite{Sakurai}. This expression can be written more compactly as
\begin{equation}\label{kernel}
    \hat{\Delta}_{q}(\vec{n})=\hat{\Delta}_{q}(\theta,\phi)=\frac{1}{2}(1_{2}+\sqrt{3}\vec{n}\cdot\vec{\sigma}),
\end{equation}
where $\vec{n}$ is the unit vector on the sphere given by $\vec{n}=(\sin\theta\cos\phi, \sin\theta\sin\phi,\cos\theta)^{T}\in\mathbb{R}^{3}$ and $1_{2}$ is the $2\times 2$ identity matrix operator.
Here, $\hat{\Pi}_{q}$ denotes the parity operator, defined as
\begin{equation}\label{Parity}
	\hat{\Pi}_q=1_{2}+\sqrt{3}\sigma_z, 
\end{equation}
This particular form of the spin-parity operator arises from analyzing the $SU(N)$ coherent states on the complex projective space, as discussed in \cite{Perelomov,Nemoto}. It is referred to as a parity operator because it encodes the antipodal symmetry of the sphere. In the spherical phase space $S^{2}$, this symmetry corresponds to the map $\vec{n}\to-\vec{n}$, which plays the role of phase–space inversion. This transformation is the natural analogue of the parity operation in the flat phase space $\mathbb{R}^{2n}$, where the canonical variables transform as 
$(q,p)\to(-q,-p)$.
The operator $\hat{U}(\phi,\theta,\Phi)$ appearing in Eq.~\ref{Delta} provides the $SU(2)$ counterpart of the displacement operator in continuous-variables systems. In this case, it corresponds to an $SU(2)$ rotation operator, expressed through a particular Euler-angle decomposition:  
\begin{equation}\label{SU2Rotation}
	\hat{U}(\phi,\theta,\Phi)=exp\left(i\hat{\sigma}_z\frac{\phi}{2}\right)exp\left(-i\hat{\sigma}_y \frac{\theta}{2}\right)exp\left(-i\hat{\sigma}_z\frac{\Phi}{2}\right).
\end{equation}
It is worth noting that the kernel $\hat{\Delta}_q(\theta,\phi)$ does not depend explicitly on the parameter $\Phi$. This independence reflects the gauge freedom inherent in the Euler angle parametrization. As a consequence, the associated phase-space functions are naturally defined on the coadjoint orbit of $SU(2)$, which in this case corresponds to the two-sphere $S^{2}$.

With the kernel (\ref{kernel}) at hand, the phase-space symbol associated with any operator $\hat{A}$ acting on the qubit Hilbert space $\mathcal{H}=\mathbb{C}^{2}$ is defined as
\begin{equation}\label{Wmap}
    W_{\hat{A}}(\vec{n}):=\tr(\hat{A}\hat{\Delta}_{q}(\vec{n})).
\end{equation}
This mapping can be inverted, allowing the operator $\hat{A}$ to be reconstructed from its phase–space symbol according to
\begin{equation}
    \hat{A}=\frac{1}{2\pi}\int_{0}^{2\pi}\int_{0}^{\pi}W_{\hat{A}}(\vec{n})\hat{\Delta}_{q}(\vec{n})\sin\theta \,d\theta d\phi.
\end{equation}
Since any operator acting on $\mathbb{C}^{2}$ can be expressed in the Pauli basis as $\hat{A}=a_{a}1_{2}+\vec{a}\cdot\vec{\sigma}$, where $a_{0}\in\mathbb{R}$ and $\vec{a}\in\mathbb{R}^{3}$, the corresponding phase-space function $W_{\hat{A}}(\vec{n})$ takes the form
\begin{equation}\label{symbol}
    W_{\hat{A}}(\vec{n})=a_{0}+\sqrt{3}\vec{a}\cdot\vec{n}.
\end{equation}

A noncommutative structure on the phase-space algebra arises naturally within the SW quantization formalism. To see this, let $\hat{A}$ and $\hat{B}$  be two arbitrary operators acting on the Hilbert space $\mathbb{C}^{2}$. Through the SW correspondence ($\ref{Wmap}$) and the definition of the symbol (\ref{symbol}), the phase-space function on the sphere associated with the product of operators, $W_{\hat{A}\hat{B}}(\vec{n})$, induces a noncommutative product between functions on the phase space, known as the star product, given by
\begin{eqnarray}\label{star}
    W_{\hat{A}\hat{B}}(\vec{n})&=&W_{\hat{A}}\star W_{\hat{B}}(\vec{n})=\tr(\hat{A}\hat{B}\hat{\Delta}(\vec{n}))\nonumber \\
    &=&(a_{0}b_{0}+\vec{a}\cdot{\vec{b}})+\sqrt{3}(a_{0}\vec{b}+b_{0}\vec{a}+i\vec{a}\times\vec{b})\cdot{\vec{n}}.
\end{eqnarray}
Hence, the corresponding star product can be interpreted as a phase-space realization of the algebra of complexified quaternions $\mathbb{H}\otimes\mathbb{C}$ \cite{Lounesto,Girard,Rodrigues}. To see this, let us introduce the complex quaternions $q_{A}=a_{0}+i\vec{a}$ and $q_{B}=b_{0}+i\vec{b}$, where $\vec{a}=a_{1}\mathbf{i}+a_{2}\mathbf{j}+a_{3}\mathbf{k}$ and similarly for $\vec{b}$. The basis elements $\mathbf{i}, \mathbf{j}, \mathbf{k}$ satisfy the quaternion relations
\begin{equation}
\mathbf{i}^{2}=\mathbf{j}^{2}=\mathbf{k}^{2}=-1, \;\; \mathbf{i}\mathbf{j}=\mathbf{k},
\end{equation}
together with cyclic permutations. With this identification, the star product of the symbols associated with the operators $\hat{A}$ and $\hat{B}$ takes the form $W_{\hat{A}}\star W_{\hat{B}}(\vec{n})=q_{A}q_{B}$. This reflects the well-known algebraic isomorphism between the space of linear operators acting con $\mathbb{C}^{2}$, namely $M_{2}(\mathbb{C})$, and the algebra of complexified quaternions, $M_{2}(\mathbb{C})\simeq \mathbb{H}\otimes\mathbb{C}$. This star product reproduces the Lie algebra structure of $\mathfrak{su}(2)$. Indeed, since $W_{\sigma_{i}}(\vec{n})=\sqrt{3}n_{i}$, the star product fulfills
\begin{equation}
    W_{\sigma_{i}}\star W_{\sigma_{j}}=\delta_{ij}+i\epsilon_{ijk}W_{\sigma_{k}},
\end{equation}
which implies, 
\begin{equation}
    \{W_{\sigma_{i}},W_{\sigma_{j}}\}_{M}:=W_{\sigma_{i}}\star W_{\sigma_{j}}-W_{\sigma_{j}}\star W_{\sigma_{i}}=2i\epsilon_{ijk}W_{\sigma_{k}}
\end{equation}
where $\{\cdot,\cdot\}_{M}$ denotes the Moyal braket. Therefore, the star product on the sphere (\ref{star}), together with the Moyal braket, provides a realization of the $\mathfrak{su}(2)$ Lie algebra at the level of phase-space functions.

\section{Star exponentials and path integrals}\label{sec:SE}

The propagator or transition amplitude 
\begin{equation}
    K(\psi_{f},t_{f};\psi_{0},t_{0})=\braket{\psi_{f},t_{f}|\psi_{0},t_{0}},
\end{equation}
encodes the full dynamical content of the quantum evolution of a system, as it corresponds to the kernel of the unitary time-evolution operator \cite{Cohen}. When the Hamiltonian operator does not depend explicitly on time, the evolution operator is given by $\hat{U}(t_{f},t_{0})=e^{-\frac{i}{\hbar}\hat{H}(t_{f}-t_{0})}$, and the propagator can be expressed as
\begin{equation}
    K(\psi_{f},t_{f};\psi_{0},t_{0})=\braket{\psi_{f}|\hat{U}(t_{t_{f}},t_{0})|\psi_{0}}.
\end{equation}
In the short time limit, it reduces to
\begin{equation}
    \lim_{t_{f}\to t_{0}} K(\psi_{f},t_{f};\psi_{0},t_{0})=\braket{\psi_{f}|\psi_{0}}
\end{equation}
which guarantees the continuity of the quantum evolution. Moreover, the value of the function $|K(\psi_{f},t_{f};\psi_{0},t_{0})|^{2}$ represents the conditional probability distribution of finding the system in the state $\ket{\psi_{f}}$ at time $t_{f}$, starting form a given initial state $\ket{\psi_{0}}$ at time $t_{0}$.

Let $\hat{\rho}_{f,0}=\ket{\psi_{f}}\bra{\psi_{0}}\in M_{2}(\mathbb{C})$ be a non diagonal density operator, where $\ket{\psi_{f}}, \ket{\psi_{0}}\in\mathbb{C}^{2}$. Using the Pauli basis, the density operator $\hat{\rho}_{f,0}$ admits the decomposition
\begin{equation}
    \hat{\rho}_{f,0}=\frac{1}{2}\braket{\psi_{f}|\psi_{0}}+\frac{1}{2}\braket{\psi_{f}|\vec{\sigma}|\psi_{0}}\cdot\vec{\sigma}.
\end{equation}
By means of the traciality condition of the SW correspondence, the propagator can be written as
\begin{eqnarray}
    K(\psi_{f},t_{f};\psi_{0},t_{0})&=&\tr(\hat{\rho}_{f,0}\hat{U}(t_{f},t_{0})) \nonumber\\
    &=&\frac{1}{2\pi}\int_{0}^{2\pi}\int_{0}^{\pi}W_{ \hat{\rho}_{f,0}}(\vec{n})W_{\hat{U}(t_{f},t_{0})}(\vec{n})\sin\theta \,d\theta d\phi,
\end{eqnarray}
where $W_{ \hat{\rho}_{f,0}}(\vec{n})$ and $W_{\hat{U}(t_{f},t_{0})}(\vec{n})$ denotes the SW symbols of the operators $\hat{\rho}_{f,0}$ and $\hat{U}(t_{f},t_{0})$ on the sphere, respectively. On the one hand, the phase-space function $W_{ \hat{\rho}_{f,0}}(\vec{n})$ corresponds to a non-diagonal Wigner function and, following Eq. (\ref{Wmap}), takes the form
\begin{equation}
    W_{ \hat{\rho}_{f,0}}(\vec{n})=\frac{1}{2}\braket{\psi_{f}|\psi_{0}}+\frac{\sqrt{3}}{2}\braket{\psi_{f}|\vec{\sigma}|\psi_{0}}\cdot\vec{n},
\end{equation}
while, on the other hand, the symbol of the unitary time evolution operator is given by the star exponential of the phase-space symbol $W_{\hat{H}}(\vec{n})$ of the Hamiltonian operator $\hat{H}$,
\begin{equation}\label{WU}
 W_{\hat{U}(t_{f},t_{0})}(\vec{n})=\Exp{\left(-\frac{i}{\hbar}(t_{f}-t_{0})W_{\hat{H}}(\vec{n})\right)},
\end{equation}
where the star exponential is defined through the series expansion
\begin{equation}\label{SE}
 \hspace{-6em}   \Exp{\left(-\frac{i}{\hbar}(t_{f}-t_{0})W_{\hat{H}}(\vec{n})\right)}=1-\frac{i}{\hbar}(t_{f}-t_{0})W_{\hat{H}}(\vec{n})+\frac{1}{2!}\left(\frac{-i(t_{f}-t_{0})}{\hbar}\right)^{2}W_{\hat{H}}(\vec{n})\star W_{\hat{H}}(\vec{n})+\cdots
\end{equation}
with the star product defined in Eq. (\ref{star}). Then, the propagator may be identified as follows
\begin{equation}\label{Prop}
\hspace{-4em}    K(\psi_{f},t_{f};\psi_{0},t_{0})=\frac{1}{2\pi}\int_{0}^{2\pi}\int_{0}^{\pi}W_{ \hat{\rho}_{f,0}}(\vec{n})\Exp{\left(-\frac{i}{\hbar}(t_{f}-t_{0})W_{\hat{H}}(\vec{n})\right)}\sin\theta \,d\theta d\phi.
\end{equation}
The former expression shows that the propagator can be formulated in terms of the star exponential of the phase-space symbol associated with the Hamiltonian, where the classical phase space is given by a sphere. This provides the natural analogue, in the $SU(2)$ setting, of the corresponding formulation of the propagator in terms of star exponentials on Euclidean phase spaces, as developed in \cite{StarExp}.

The propagator for qubits also admits a path integral representation in terms of $SU(2)$ coherent states, leading to a functional integral over $S^{2}\simeq \mathbb{CP}^{1}$ \cite{SU(2)path}. Any qubit sate can be parametrized as
\begin{equation}
    \ket{\zeta}=\cos\frac{\theta}{2}\ket{0}+e^{i\phi}\sin\frac{\theta}{2}\ket{1},
\end{equation}
which corresponds to an $SU(2)$ coherent state for a spin-1/2 system. These states satisfy the resolution of the identity 
\begin{equation}
    \frac{1}{2\pi}\int_{0}^{2\pi}\int_{0}^{\pi}\ket{\zeta}\bra{\zeta}\sin\theta \,d\theta d\phi=1_{2}.
\end{equation}
By inserting this resolution of the identity and performing a standard time-slicing procedure for the evolution operator $\hat{U}(t_{f},t_{0})$, the transition amplitude $K(\psi_{f},t_{f};\psi_{0},t_{0})$ can be written as a functional integral over continuous paths on $S^{2}$,
\begin{equation}\label{path}
    K(\psi_{f},t_{f};\psi_{0},t_{0})=\int_{\zeta_{0}}^{\zeta_{f}}\mathcal{D}\mu(\zeta)\exp\left(\frac{i}{\hbar}S[\zeta]\right),
\end{equation}
where $\ket{\psi_{0}}=\ket{\zeta_{0}}$ and $\ket{\psi_{f}}=\ket{\zeta_{f}}$. The path integral measure is given by
\begin{equation}
    \mathcal{D}\mu(\zeta)=\lim_{N\to\infty}\prod_{k=1}^{N-1}\frac{1}{2\pi}\sin\theta_{k}d\theta_{k}d\phi_{k}.
\end{equation}
The action functional takes the form
\begin{equation}
    S[\zeta]=\int_{t_{0}}^{t_{f}}\left[i\hbar\braket{\zeta|\dot{\zeta}}-H_{B}(\zeta)\right]dt.
\end{equation}
where the first term represents the geometric Berry phase and defines a symplectic potential whose exterior derivative yields the Kirillov–Kostant–Souriau two-form (\ref{wKKS}), endowing $S^{2}$ with its symplectic structure. The function $H_{B}(\zeta)=\braket{\zeta|\hat{H}|\zeta}$ corresponds to the Berezin symbol of the Hamiltonian operator $\hat{H}$, and thus plays the role of the classical Hamiltonian governing the dynamics on the sphere.

The above expressions (\ref{Prop}) and (\ref{path}) show that the propagator admits two equivalent yet conceptually distinct representations, the latter corresponding to a geometric formulation in terms of path integrals over trajectories on the sphere, while the former stands for an algebraic formulation in terms of star exponentials within the deformation quantization framework. This duality is directly analogous to the situation in Euclidean phase space, where the propagator can be described either through Feynman path integrals on $\mathbb{R}^{2n}$ or by Moyal star exponentials of the Hamiltonian symbol \cite{StarExp, FeynmanKac,SETDHO,Fermi}. In the present case, however, both constructions are naturally defined on the curved phase space given by the sphere, reflecting the underlying coadjoint orbit structure of the system.

\subsection{Examples}
In this subsection, we present illustrative examples that clarify the structure of the star product, together with a concrete application to the dynamics of a two-level system. In particular, we show how the star product naturally encodes the system’s dynamical evolution in phase space.

\subsubsection{Two level system in a magnetic field.}

As an illustrative example, consider a spin-1/2 system in an external magnetic field, described by the Hamiltonian $\hat{H}=-(\gamma/2)\vec{B}\cdot\vec{\sigma}$, where $\gamma$ denotes the gyromagnetic ratio and $\vec{B}$ is the magnetic field. This system provides a model for two-level quantum dynamics, describing the interaction between a spin degree of freedom and an external field \cite{Cohen}. As a particular but not trivial example, let us choose $\ket{\psi_{0}}=\ket{\uparrow_{z}}$ and $\ket{\psi_{f}}=\ket{\downarrow_{z}}$, and take the magnetic field along the $x$-direction, $\vec{B}=B\hat{x}$. The Hamiltonian then reads $\hat{H}=-(\gamma B/2)\sigma_{x}$. Using Eq. (\ref{symbol}) its associated phase-space symbol is given by
\begin{equation}
    W_{\hat{H}}(\vec{n})=-\frac{\sqrt{3}\gamma B}{2}n_{x}.
\end{equation}
Moreover, the corresponding classical dynamics is governed by spin precession, thereby illustrating the consistency between the quantum evolution and the underlying symplectic structure of the sphere. Therefeore, by combining Eq. (\ref{WU}) with the definition of the star exponential in Eq. (\ref{SE}), the phase-space symbol of the unitary time evolution operator can be written as
\begin{equation}
    W_{\hat{U}(t_{f},t_{0})}(\vec{n})=\cos\left(\frac{\gamma B t}{2\hbar}\right)+i\sqrt{3}\sin\left(\frac{\gamma B t}{2\hbar}\right)n_{x},
\end{equation}
where $t_{0}=0$ and $t_{f}=t$. The Wigner function associated with the density matrix  $\hat{\rho}_{f,0}=\ket{\downarrow_{z}}\bra{\uparrow_{z}}$ is
\begin{equation}
    W_{\hat{\rho}_{f,0}}(\vec{n})=\frac{\sqrt{3}}{2}(n_{x}+in_{y}).
\end{equation}
Thus, the propagator (\ref{Prop}) takes the form
\begin{equation}
    K(\downarrow_{z},t; \uparrow_{z},0)=i\sin\left(\frac{\gamma B t}{2\hbar}\right).
\end{equation}
Similarly, one finds
\begin{equation}
    K(\uparrow_{z},t; \uparrow_{z},0)=\cos\left(\frac{\gamma B t}{2\hbar}\right).
\end{equation}
Their corresponding transition probabilities then read as
\begin{equation}
     |K(\downarrow_{z},t; \uparrow_{z},0)|^{2}=\sin^{2}\left(\frac{\gamma B t}{2\hbar}\right), \;\; |K(\uparrow_{z},t; \uparrow_{z},0)|^{2}=\cos^{2}\left(\frac{\gamma B t}{2\hbar}\right).
\end{equation}
These expressions describe the coherent oscillatory exchange of population between the two levels, characteristic of Rabi dynamics in a driven two-level system \cite{Knight}. The oscillation frequency is set by the coupling $\gamma B/\hbar$, reflecting the strength of the interaction with the external field.

\subsubsection{Poisson structure induced by the star product.} The Poisson structure associated with the star product (\ref{star}) can be obtained by considering its antisymmetric part, namely, the Moyal bracket, and subsequently identifying its classical limit. The Moyal bracket between the phase-space symbols $W_{\hat{A}}$ and $W_{\hat{B}}$ is defined as
\begin{equation}
    \{W_{\hat{A}},W_{\hat{B}}\}_{M}:=W_{\hat{A}}\star W_{\hat{B}}-W_{\hat{B}}\star W_{\hat{A}}=2\sqrt{3}i(\vec{a}\times\vec{b})\cdot{\vec{n}}.
\end{equation}
This relation induces, by bilinearity and the Leibniz rule, a bracket on the algebra of smooth functions $C^{\infty}(S^{2})$. In particular, evaluating the Moyal bracket on the coordinate functions $n_{i}$ yields
\begin{equation}
    \{n_{i},n_{j}\}=\frac{2i}{\sqrt{3}}\epsilon_{ijk}n_{k},
\end{equation}
from which one defines the corresponding Poisson bracket by removing the factor of $i$, that is, $ \{n_{i},n_{j}\}=\frac{2}{\sqrt{3}}\epsilon_{ijk}n_{k}$. Extending this bracket to arbitrary function $f,g\in C^{\infty}(S^{2})$ by derivation, obe obtains
\begin{equation}
    \{f,g\}(\vec{n})=\sum_{i,j}^{3}\frac{\partial f}{\partial n_{i}}\frac{\partial g}{\partial n_{j}}\{n_{i},n_{j}\}=\sum_{i,j,k}^{3}\frac{2}{\sqrt{3}}\epsilon_{ijk}n_{k}\frac{\partial f}{\partial n_{i}}\frac{\partial g}{\partial n_{j}}.
\end{equation}
This expression can be written compactly as
\begin{equation}
    \{f,g\}(\vec{n})=\frac{2}{\sqrt{3}}\vec{n}\cdot \left(\nabla_{\vec{n}}f\times \nabla_{\vec{n}}g\right).
\end{equation}
This structure is not arbitrary, but arises geometrically from the symplectic form on the sphere. Indeed, in spherical coordinates, the bracket takes the form 
\begin{equation}
     \{f,g\}(\vec{n})=\frac{2}{\sqrt{3}}\sin\theta\left(\frac{\partial f}{\partial\theta}\frac{\partial g}{\partial\phi}-\frac{\partial g}{\partial\theta}\frac{\partial f}{\partial\phi}\right)
\end{equation}
which is precisely, up to a constant, the inverse of the symplectic form of the sphere
\begin{equation}
    \omega=\sin\theta d\theta\wedge d\phi.
\end{equation}
Consequently, the Poisson bracket is the one induced by the KKS symplectic structure on the coadjoint orbit $S^{2}$, thereby establishing the equivalence between the algebraic structure derived from the star product and the intrinsic symplectic geometry of the sphere. It is worth emphasizing that no explicit factor of $\hbar$ appears in the Poisson structure induced by the star product. This is a consequence of working with dimensionless phase-space variables where $\hbar$ is effectively absorbed into the definition of the spin operators and the radius of the coadjoint orbit. From the perspective of geometric quantization, the deformation parameter is effectively given by the inverse spin $1/j$. In the qubit case, $j=\frac{1}{2}$, the star product is already exact, and no intrinsic semiclassical limit ($\hbar\to 0$) is available within the system, Instead, a semiclassical regime emerges only in the large-$j$ limit, where an expansion in powers of $1/j$ becomes meaningful \cite{Woodhouse,GQ}.

\section{Conclusions}\label{sec:Conclusions}

In this work, we have developed a phase-space formulation of qubit systems based on the geometry of coadjoint orbits and the Stratonovich–Weyl correspondence, providing an explicit realization of deformation quantization on the sphere. In this framework, the star product induces a noncommutative algebra on $C^{\infty}(S^{2})$ which faithfully reproduces the operator algebra of $M_{2}(\mathbb{C})$, and can be identified with the algebra of complexified quaternions. Moreover, its antisymmetric part encodes the Lie–Poisson structure associated with the Kirillov–Kostant–Souriau symplectic form. Furthermore, we have shown that quantum dynamics can be described entirely in phase space, either through star exponentials of the Hamiltonian symbol or via coherent-state path integrals, establishing a precise correspondence between algebraic and geometric formulations of the propagator. The results obtained here naturally suggest several directions for further investigation. A particularly relevant extension concerns the generalization to multipartite systems. For a system of $n$ qubits, the underlying symmetry group becomes $SU(N)$, and the associated classical phase spaces are no longer spheres, but higher-dimensional coadjoint orbits which are diffeomorphic to flag manifolds \cite{Flags}. These manifolds can be described as homogeneous spaces of the form
\begin{equation}
    \mathbb{F}\simeq SU(N)/U(n_{1})\times\cdots \times U(n_{k}),
\end{equation}
where $(n_{1},\ldots, n_{k})$ defines a partition of $N$. Geometrically, flag manifolds parametrize nested sequences of subspaces in $\mathbb{C}^{N}$, and thus provide a natural generalization of projective spaces and spheres. As coadjoint orbits, they inherit canonical symplectic structures given by the KKS form, together with compatible complex and Riemannian structures, making them K\"ahler manifolds. From the perspective of deformation quantization, these spaces provide a rich arena in which noncommutative structures can be constructed beyond the $SU(2)$ case. In particular, the associated star products are expected to reflect the complexity of multipartite operator algebras, while their semiclassical limits retain the Lie structure of $\mathfrak{su}^{*}(N)$. In this context, an important open problem is to determine whether these geometric structures can be exploited to intrinsically characterize quantum features such as entanglement, quantum correlations, and nonclassicality. Flag manifolds, as phase spaces of composite systems, provide a natural geometric framework for describing the global structure of the Hilbert space and its stratification into submanifolds, including the separable and entangled sectors. This suggests the possibility of identifying geometric invariants, symplectic structures, or curvature properties associated with entanglement measures. Moreover, the interplay between star products and multipartite structures may provide new insights into how nonlocal correlations emerge from the underlying noncommutative algebra.

In particular, we aim to develop the construction of Stratonovich–Weyl correspondences and star products on these higher-dimensional coadjoint orbits, as well as their associated path integral formulations. A central objective is to assess whether the phase-space geometry of flag manifolds offers a natural and computationally efficient framework for the geometric characterization of entanglement and other features in multi-qubit systems. Some attempts in this direction will be reported in a future communication.

\section*{Acknowledgments}
  The authors would like to acknowledge support from SNII CONAHCYT-Mexico. JBM acknowledge financial support from Marcos Moshinsky foundation.  AM acknowledges financial support from COPOCYT under project 2467 HCDC/2024/SE-02/16 (Convocatoria 2024-03, Fideicomiso 23871). MSC acknowledge financial support from SECIHTI under project Estancias Posdoctorales por México 2023(1).

\section*{Data Availability Statement}
The authors declare that no data were generated or analyzed in the course of this research.

\bibliographystyle{unsrt}

\end{document}